# Solid State Transformer – Next Generation MV Interface for Native DC loads

Steffan Hansen, Marco Liserre, Mohamed Moharram

**Summary:**

Electrification is increasing electricity demand across transport, industry, data centers and infrastructure [3]. At the same time, grid expansion is struggling to keep pace, resulting in long lead times and limited connection capacity for new high-power applications [4].

Battery energy storage systems can help overcome these constraints by reducing peak grid demand and increasing the utilization of existing connections [2]. However, batteries, electric vehicles, photovoltaic systems and many modern industrial loads are fundamentally DC-based or rely on internal DC links [5].

Conventional electrical architectures connect these assets through a low-frequency transformer and several separate converters. This results in repeated power conversion, additional losses, duplicated hardware and complex system integration [6].

Solid state transformers (SST) offer an alternative. They combine medium-voltage conversion, galvanic isolation, digital control and AC or DC output within one actively controlled platform [7].

Many SST concepts are optimized around one principal application or output voltage. PowerNexa is developing a modular multi-port SST architecture designed to integrate battery storage, charging systems and other DC loads operating at different voltage levels within one medium-voltage grid interface.

The objective is not simply to replace the transformer. It is to reduce the total number of conversion stages, controllers and system interfaces required for today's energy applications [8].

**1. The Grid Is Reaching Its Limits**

The electrical grid was designed around centralized generation, one-directional power flow and predominantly AC-based consumption. That model is changing. Electric mobility, industrial electrification, data centers, heating systems and renewable generation are increasing demand at distribution-grid level [3]. Many new applications require several megawatts of connection capacity, while grid reinforcement can involve long planning, permitting and construction periods. As a result, customers may face limited connection capacity, costly grid upgrades, long project delays and restrictions on operating power [4].

Battery storage can reduce these constraints by supplying power during demand peaks and charging when grid utilization is lower. This allows the available grid connection to be used more efficiently. For many high-power sites, battery storage is therefore becoming part of the core electrical infrastructure [2].

**2. The Energy System Is Becoming More DC-Based**

A growing number of energy assets are internally DC-based or depend on power-electronic DC links such as [5]:

- battery energy storage
- electric vehicles and truck-charging
- data-center power systems
- variable-speed industrial drives
- converter-based renewable generation

### About PowerNexa

PowerNexa is the project name for a start- up initiative driven by Marco Liserre and Steffan Hansen with the ambition to develop the next generation power grid interface.

### Battery integration and DC loads

Ongoing electrification is fundamentally changing how electrical energy is produced and consumed. DC loads are becoming an integral part of the infrastructure while battery integration is increasingly needed to manage peak demand and address grid capacity constraints [2].

### The next generation grid interface

PowerNexa aims to combine a fully controllable Medium Voltage grid interface with native battery integration and multiple DC voltage levels, based on an architecture designed for high efficiency, modularity and redundancy.

Conventional distribution systems still connect these assets through AC networks. A battery-buffered charging hub for instance, may therefore require [9]:

1. a medium-voltage transformer
2. a battery PCS
3. a charger rectifier
4. vehicle-level DC/DC converters
5. separate control and protection systems

Each conversion stage adds cost, losses, cooling requirements, footprint and control complexity [6]. The key question is therefore: Can medium-voltage electricity be converted and distributed more directly to batteries and DC loads?

## 3. What a Solid-State Transformer Changes

A conventional transformer changes voltage through passive electromagnetic conversion at grid frequency. A solid-state transformer adds actively controlled power electronics and provides galvanic isolation through a medium frequency transformer [7]. As a result, an SST has the potential of becoming not only smarter but also significantly smaller in size and weight and provide lower total losses [10].

Some of the smart features an SST potentially can provide are [7]:

- power-factor control
- active and reactive power regulation
- voltage and frequency support
- direct DC output
- battery integration
- monitoring and communication
- and coordinated energy management

The principal value of an SST is therefore not only voltage transformation. It is the integration of transformation, conversion and control within one grid interface.

## 4. SST Architectures

The term SST covers a wide band of different topologies for different applications. Not all of above-mentioned advantages are obtained with all possible topologies and configurations. Therefor this section will cover some of the most significant topology choices.

### Classification by Conversion Stages

Typically, the classification of SST is done based on number of conversion stages and modularity. The number of conversion stages has a significant influence on the capabilities of the SST. While a higher number of conversion stages are expected to have higher controllability and flexibility while lower number of conversion stages promises lower cost and potential higher efficiency.

A single-stage SST converts power directly without a separate intermediate DC link. It can offer compact architecture and low component count, but provides limited decoupling between input and output, hence a single-stage SST will have significant limitations with regards of input and output controllability and thus limitations in voltage quality and smart grid features. A single-stage SST can in principle provide either DC or AC output voltage.

A two-stage SST includes one DC link, either on the medium-voltage or low-voltage side. This creates one direct DC connection point and improves controllability. Input and output voltage quality is significantly improved over a single stage SST, and smart grid features are possible depending of MV converter side topology. As the single-stage SST, the two-stage SST can be configured to provide AC or DC voltage output. If DC output is required a two-stage SST is the preferred choice.

For completion, the three-stage SST should be mentioned here as well. The three-stage SST is mainly relevant when AC output is required. A  three-stage SST contains DC links on both sides of the medium-frequency transformer.

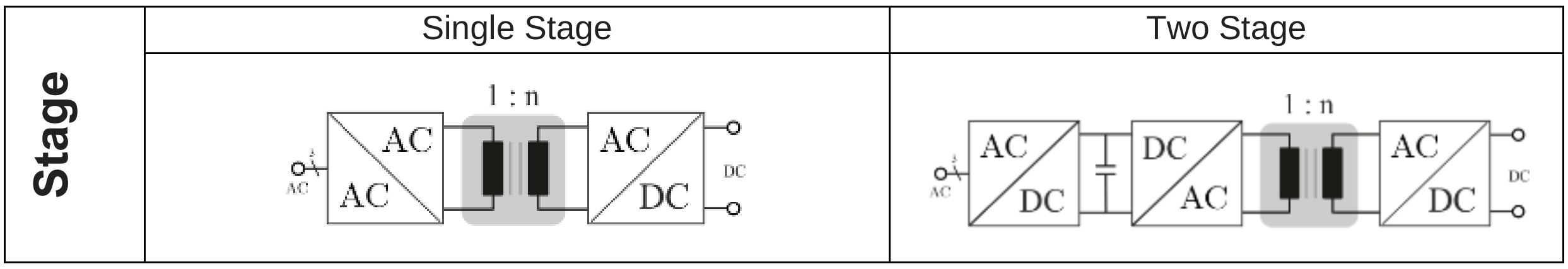

**Figure 1:** Principal diagrams of a single- and two-stage topology

**Classification by Modularity**

SST architecture can be classified not only by the number of conversion stages, but also by the level at which modularity is introduced. In General modularity can be divided into

- **Non-modular:** the converter is designed as one integrated power stage.
- **Modular:** the system is assembled from repeated converter or transformer modules.

Each modular element can then be obtained by three levels of integration:

- **At cell level,** multiple identical power-electronic cells are connected in series or parallel to achieve the required voltage and power rating.
- **At converter level**, complete isolated converter modules are combined to increase power capacity. Each module can include: semiconductor switches, a medium-frequency transformer, local control, and an isolated output stage.
- **At system level**, complete functional systems are combined as standardized building blocks.

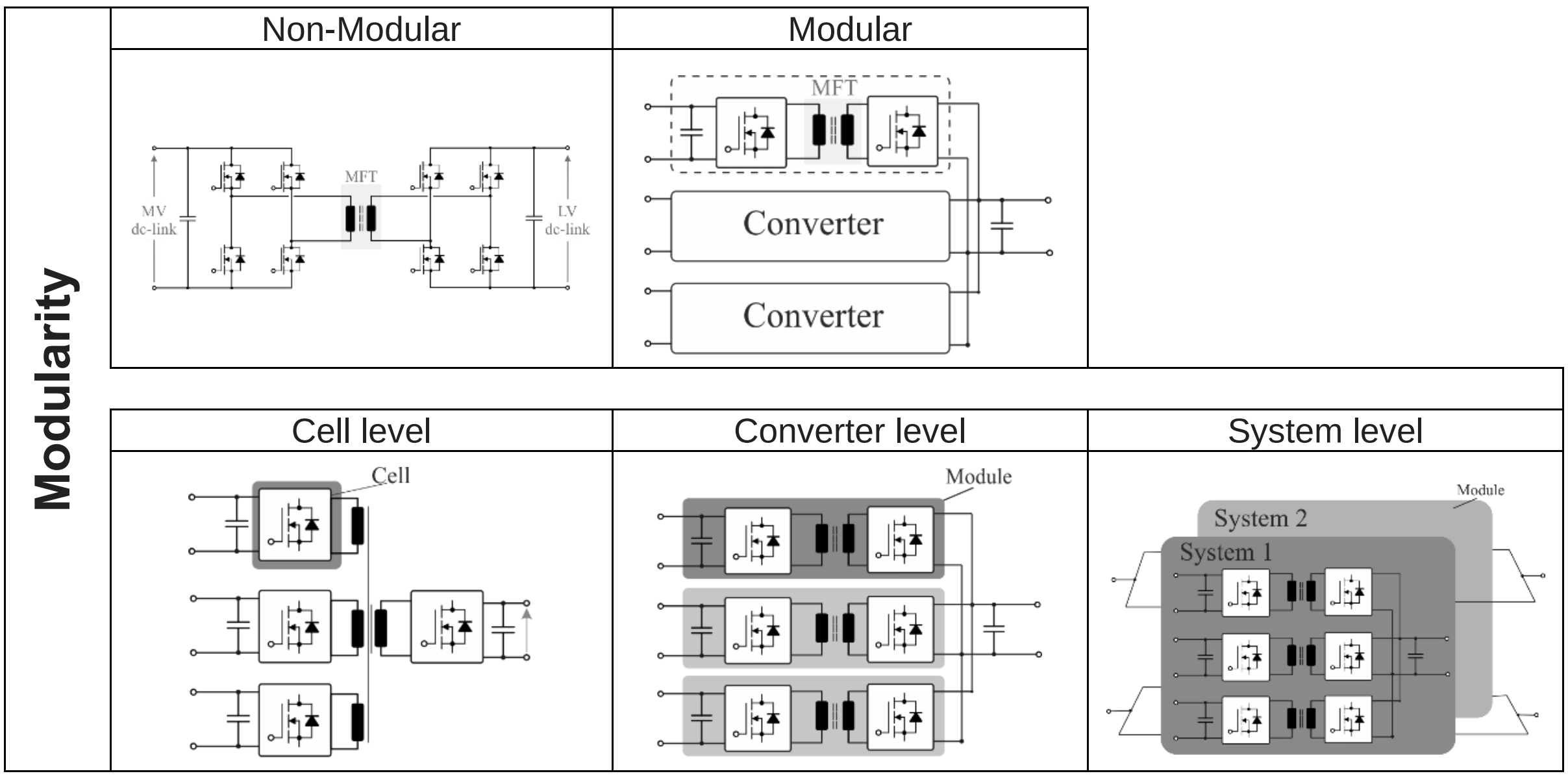

**Figure 2:** Principle diagrams of different levels of Modularity

Modularity is considered advantageous because it allows the SST to be scaled, configured and serviced using standardized building blocks. Power ratings and output functions can be adapted by adding or combining modules instead of redesigning the complete system. It can also improve availability by enabling faulty modules to be isolated or replaced while the remaining system continues operating. The highest level of modularity is obtained on a cell level.

### Why the right building block matter

As discussed above, the term **SST** encompasses a broad range of converter architectures, comprising different topologies and basic cells. These architectures may support unidirectional or bidirectional power flow and provide single or multiple DC connections. The central question is therefore which building block can best accommodate the diverse configurations required by different applications and their evolving requirements, including DC voltage level and polarity, power rating, power-flow directionality, and energy-storage integration. Identifying an optimal building block requires a systematic examination of these requirements and the associated design trade-offs.

### MV Stage topology

Special consideration should be given to the medium voltage conversion stage as the topology choice of all subsequent conversion stages highly depends on the choice of the MV conversion stage. As a modular design is preferred as described in the previous section, the choice of the MV stage is in practice restricted to two main topologies: The Modular Multilevel Converter (MMC) and Cascaded H-Bridge.

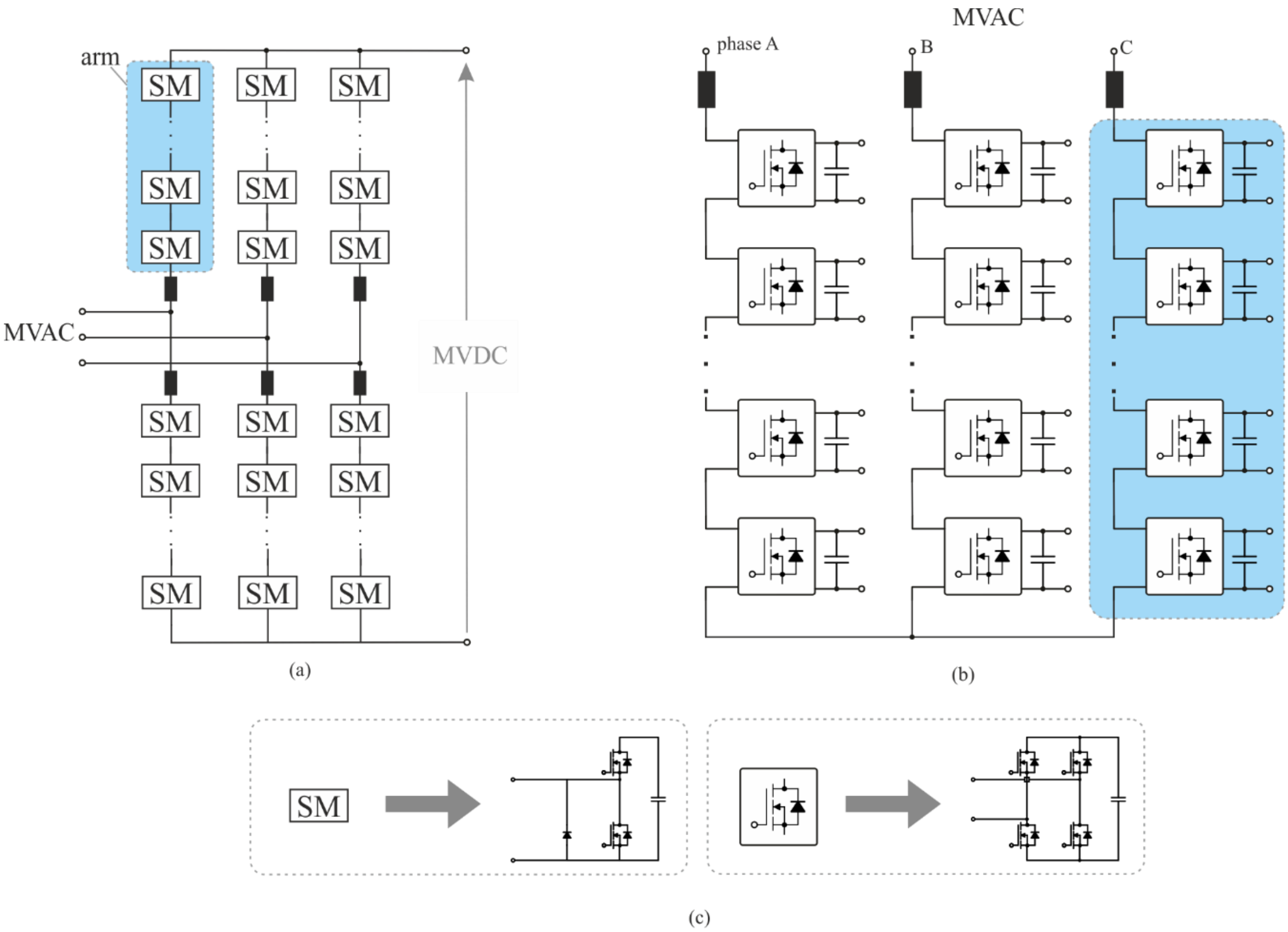

Figure 3: (a) Modular Multilevel Converter, (b) Cascaded H-Bridge, (c) Topology used for the MMC cell/modules vs CHB cell/modules.

The MMC topology is an established solution in high-voltage applications such as HVDC transmission systems, whereas the CHB topology is widely used in industrial medium-voltage applications. A key advantage of the MMC is its accessible MV DC link, which can connect to other MVDC assets. This becomes particularly relevant when an SST enables a DC-based distribution system beyond the replacement of a conventional low-frequency transformer. The CHB on the other hand has no MV DC-link but has significant advantages in terms of less complexity (ease of control), total semiconductors used, number and size of AC filter and capacitance used leading to an overall lower cost. In conventional MMC and CHB architectures, requiring an accessible MV DC link therefore favors the MMC.

As described in section 5 the PowerNexa architecture changes this, as its architecture enables a synthesized MVDC voltage while retaining a CHB-based medium-voltage front end.

### Multi-port SST

Another relevant classification of SST topologies is to consider the number of output ports. This is simple, as it is either single output or multiple output. Thus, a multi-port SST provides several output ports or voltage domains. This makes it suitable for sites combining:

- battery storage
- vehicle charging
- renewable generation
- industrial DC loads
- and optional AC distribution

One important aspect though is how the multiple port is achieved. A typical requirement would be that the ports are galvanic isolated to ensure that each port can be controlled individually. In case of a cell or module-based topology this can be achieved by modularizing the transformer as well (such an arrangement would be called double active bridge (DAB)) and group one part of the cells for one DC voltage level (e.g., 800VDC) and another group of cells for another voltage level (e.g., 1500VDC) as shown in Figure 4(a). However, this would have the drawback that each port is limited to a portion of the total SST power and that the power path between the loads always must go through the primary circuit which reduces the efficiency. Hence the DAB is not optimal for providing multiple DC outputs.

A typical configuration for a DAB is to have all cells on the output side in parallel and only provide one DC port for a common DC distribution. This is considered the reference design for SST applied targeted for data center application. Here integration of a battery can only be achieved by applying an additional DC/DC converter for controlling the battery as shown in Figure 4(b). This has of course a negative impact on overall efficiency.

A more elegant way is to solve this by applying a muti-winding transformer (MWT) which allows multiple output levels as well as providing a direct path between the ports as shown in Figure 4(c). A configuration utilizing a 3-winding Transformer is called a Trible Active Bridge (TAB)

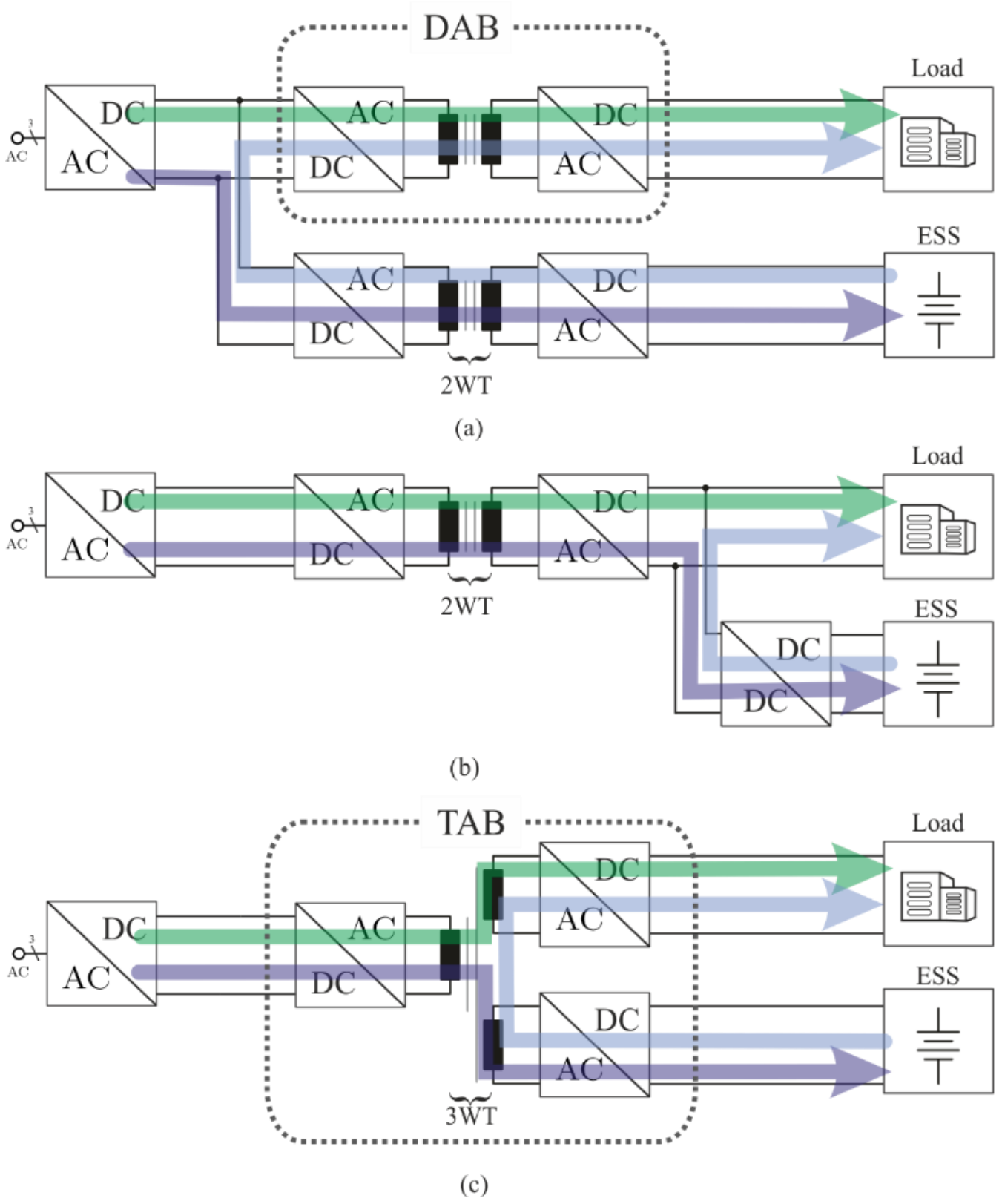

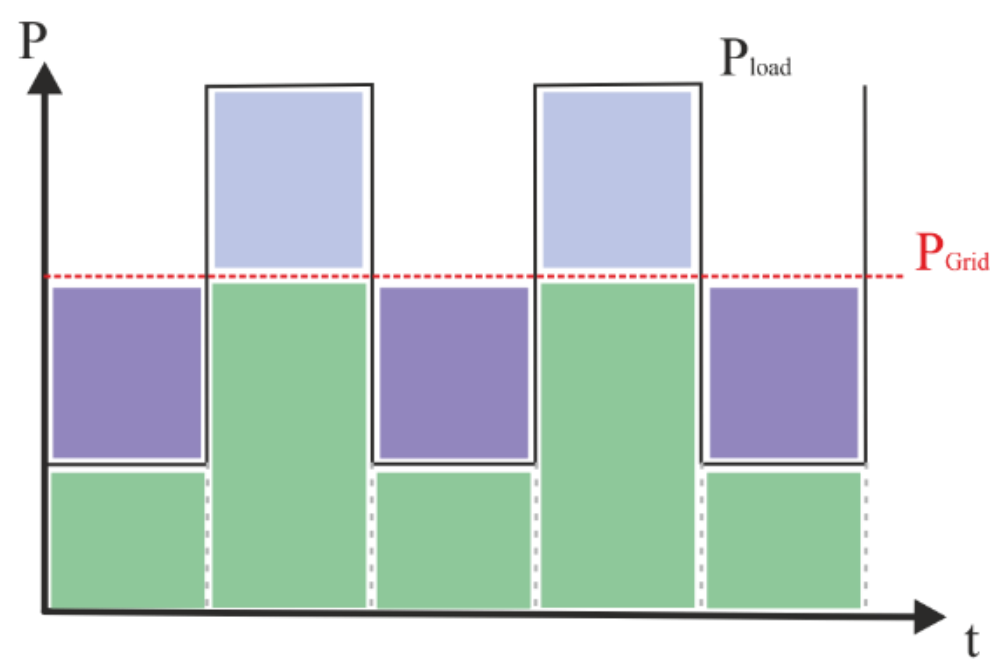

| Efficiency / Topology | Direct Power -flow | Charging BESS | BESS to Load |
|---|---|---|---|
| Multiple 2WT | $\eta_{DAB}$ 98.5% | $\eta_{DAB}$ 98.5% | $\eta_{DAB} * \eta_{DAB}$ 97.02% |
| Single 2WT +DC/DC | $\eta_{DAB}$ 98.5% | $\eta_{DAB} * \eta_{DC/DC}$ 96.7% | $\eta_{DC/DC}$ 98.2% |
| Single 3WT | $\eta_{TAB}$ 98.5% | $\eta_{TAB}$ 98.5% | $\eta_{TAB}$ 98.5% |

**Figure 4: Power flow of multiple transformers (DAB) vs a three-winding Transformer**

In a situation where the BESS is used for peak shaving the difference between the above-mentioned concepts in terms of efficiency is significant. Assuming 98.5% efficiency for the DAB [11] and TAB [12] configuration and 98.2 for the separate DC/DC converter for the battery control [13]. the overall efficiency of the system while charging the BESS 96.7% when configured as shown in Figure 4(b) vs. 98.5% for a 3-winding transformer and the DAB with multiple outputs levels. In case of providing power from the BESS the efficiency of the DAB configuration with multiple outputs and extra DC/DC converter has an efficiency of 97% and 98.2% respectively while the TAB remains at 98.5%.

## 5. The PowerNexa Concept: the key building block as universal enabler for a SST-platform

PowerNexa is developing a modular multi-port SST platform ready for several dc-voltages and integration of dc-storages by using a classical and well proven CHB-based medium-voltage front-end (also known as single-star MMC). The enabling building block is a TAB (also a well proven technology), which can be connected in several ways according to the patent [1]. The game-changing innovation lies in the flexibility of how the cells can be interconnected - and retain the potential of a single-stage variant (Figure 5a) as well as “partly” unidirectional operation (Figure 5b), both increasing further efficiency and in general guaranteeing:

- multiple controlled DC outputs
- extending dc-voltage without redesign needed
- native battery integration in the dc/dc conversion, with efficiency advantages shown in Fig. 4
- right modularity level of the power electronics

The core of the invention presented in [1] is the series or parallel interconnection of corresponding cells **across _multiple_ TABs**. This arrangement ensures that the additional port is supplied by the entire SST rather than by only a portion of it (i.e. the CHB is always balanced !). Achieving the same functionality by interconnecting DABs or adding secondary cells to a multi-winding topology would increase the number of cells and the system complexity, requiring substantial design changes. In contrast, the approach in [1] uses the same building block to extend the voltage range, provide multiple DC voltage levels, and/or integrate a battery without an additional power converter. The **TAB**, comprising only three cells, is the simplest multi-winding topology and thus offers maximum flexibility with minimum complexity.

***One TAB -> a flexible SST Platform***

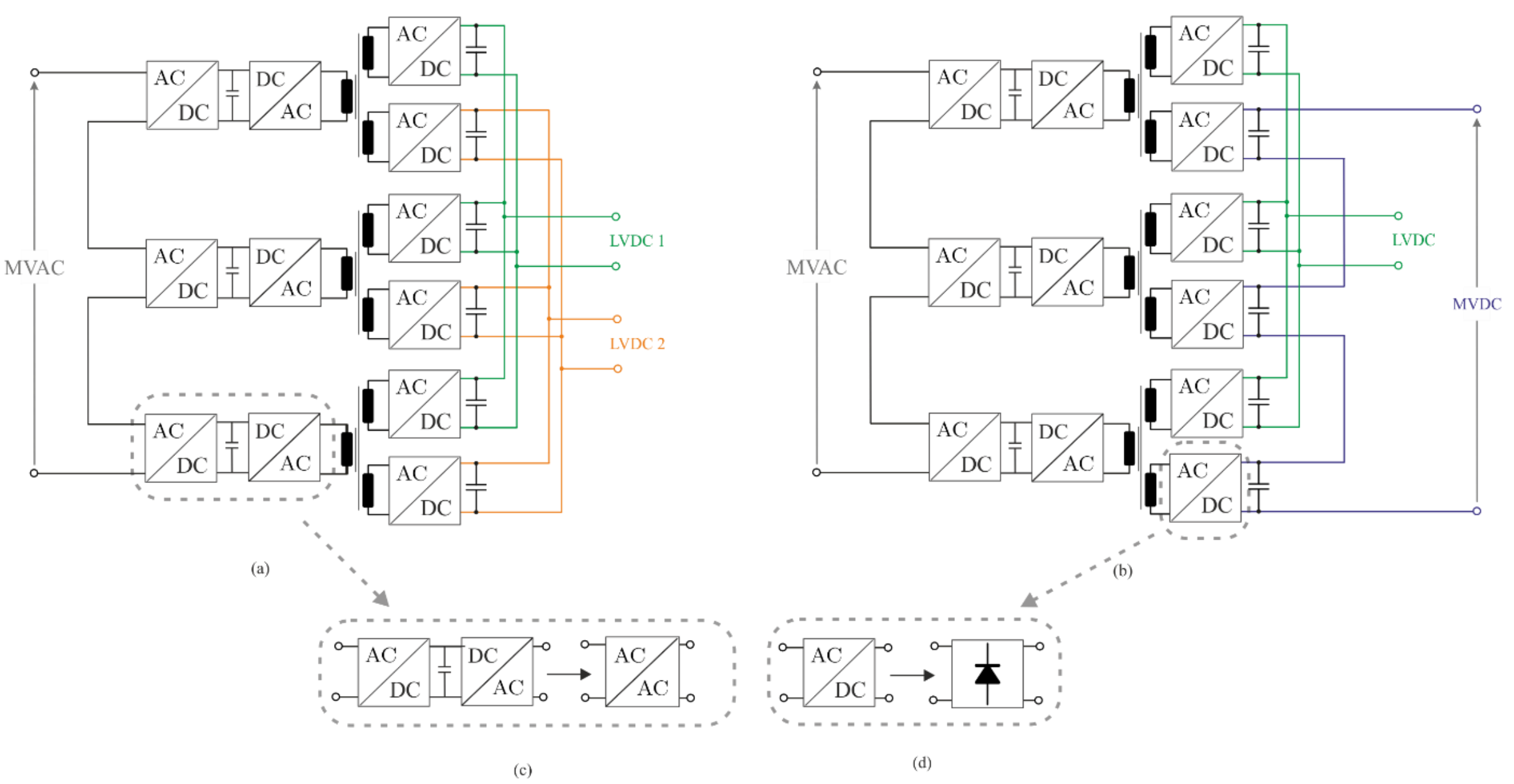

**Figure 5: PowerNexa multiport SST solution (a) Two LVDC output ports (b) LVDC and MVDC output ports (c) Combined option with AC/AC frontend (d) Low-cost option with rectifier backend**

PowerNexa's architecture is designed as a flexible SST platform for use across multiple applications, offering the following system-level benefits beyond the single-application focus of many emerging SST solutions being developed by major industry players.

- **Fewer external conversion stages:** Battery and charging integration can potentially be handled within one coordinated conversion platform.
- **Native battery integration:** The battery becomes a direct system port rather than a separate AC-connected subsystem.
- **Multiple DC voltage domains:** Different applications can operate at their required voltage levels without each requiring a complete independent grid-conversion chain.
- **Integrated control:** Grid power, battery charging, charger demand and DC-bus operation can be coordinated through one control platform associated to the dc/dc converter.
- **Modular scalability:** MMC cells and modular converter blocks can support different power ratings and provide redundancy.
- **Reduced footprint:** Shared conversion, control and cooling systems may reduce the total installed footprint compared with several separate converter cabinets.

### 6. Conclusion

The electrification of transport, industry and digital infrastructure is increasing pressure on the distribution grid. At the same time, batteries and DC-native loads are becoming central elements of modern energy systems. Conventional architecture can integrate these assets but often require several independent conversion systems. Solid-state transformers provide a path toward a more actively controlled and DC-capable grid interface.

PowerNexa aims to extend this concept through a modular multi-port architecture that integrates battery storage, charging infrastructure and several DC voltage domains within one medium-voltage platform. The commercial opportunity does not come from replacing the transformer alone. It comes from simplifying the complete system.